\documentclass[reprint,aip,showpacs,colors]{revtex4-1}
\usepackage{epsfig,color}
\usepackage{graphicx}
\usepackage{amsmath}

\begin{document}
\title{Spin current source based on a quantum point contact with local spin-orbit interaction}

\author{M. P. Nowak}
\affiliation{AGH University of Science and Technology, Faculty of Physics and Applied Computer Science,\\
al. Mickiewicza 30, 30-059 Krak\'ow, Poland}
\author{B. Szafran}
\affiliation{AGH University of Science and Technology, Faculty of Physics and Applied Computer Science,\\
al. Mickiewicza 30, 30-059 Krak\'ow, Poland}

\date{\today}


\begin{abstract}
Proposal for construction of a source of spin-polarized current based on quantum point contact with local spin-orbit interaction is presented.
We show that spin-orbit interaction present within the narrowing acts like a spin filter.
 The spin polarization of the current is discussed as a function of the Fermi energy and the width of the quantum point contact.
\end{abstract}
\maketitle

A source of spin currents that would operate in the absence of external magnetic field is considered a necessary prerequisite for practical implementation of spintronic devices.\cite{sfet}
For two-dimensional electron gas the polarized currents can be obtained using ferromagnetic contacts,\cite{hammar} but the degree of spin polarization
is typically very low.\cite{schmidt} An alternative option for generation of spin densities and spin-polarized currents without the
external magnetic fields or ferromagnets is provided\cite{debray} by the spin-orbit (SO) interaction that translates external electric fields into an effective magnetic field for moving electrons.
With the SO interaction a purely electrical control of the spin \cite{nowack,nadj,sfet} and spin filtering\cite{foldi,kalaman,molod,bednarek,nowak} is possible in particular in InGaAs that provide strong SO coupling\cite{sato} and allow for long spin diffusion lengths in quantum wires even at few Kelvin temperatures.\cite{holl}



The spin polarization\cite{rokh} induced by the Rashba\cite{rashba} coupling (that results from the electric field in the growth\cite{kim} or in-plane\cite{ngo,das} directions) is considered\cite{micolich,wang,rejec} responsible for the appearance of a shoulder-like feature\cite{thomas} for $G<2e^2/h$ in the QPC conductance.
 The possibility of spin polarization of an electron passing QPC has been considered in terms of the ratio of Rashba strength and Fermi wavevector\cite{meto} and weak spin polarization as a function of saddle-point QPC height has been studied.\cite{sablikov} Also the spin precession in SO-coupled wire has been discussed in terms of conductance of spin polarized electron.\cite{mireles}
In this paper we demonstrate the spin polarizing proprieties of a QPC with local SO interaction. The system acts like a spin current source upon application of a bias. The output spin current polarization is found for various Fermi energies and QPC widths and results from spin polarization of positive current modes within the constriction due to the SO-induced subband mixing.\cite{mroz}

We consider the system described by the Hamiltonian $H=\left[\hbar^2\textbf{k}^2/2m^*+V(\textbf{r})\right]\textbf{1} + H_{SIA}$,
where $\hbar \mathbf{k}=-i\hbar \nabla$ is the momentum operator, $\textbf{1}$ is the identity matrix and $V(\textbf{r})=V_{\mathrm{C}}(x,y)+|e|F_zzf(y)$ stands for the potential energy.
 We adopt the usual assumption that the confinement within the growth $z$ direction is strong enough to freeze all the electron wave functions in the ground-state
 of the vertical quantization. The electric field in the growth direction is applied in the region $y\in (y_1,y_2)$ and is modulated by $f(y)$ shape function,
$f(y)=(\arctan(y-y_1) + \pi/2)(-\arctan(y-y_2)+\pi/2)/\pi^2$. For $y<y_1$ and $y>y_2$ the potential gradient and hence the SO interaction vanishes.
The constriction forming the quantum point contact is defined within the $(y_1,y_2)$ segment and is modeled by potential $V_{\mathrm{C}}(x,y) = \exp\left\{ -\left[(y-Y_{\mathrm{qpc}})/(\sqrt{2}l)\right]^2 \right\}m^*\omega^2x^2/2$ where $Y_{\mathrm{qpc}}$ is the center of the QPC,  $l$ is the length, and $\hbar \omega$ defines the energy of the transverse parabolic confinement.

\begin{figure}[ht!]
\epsfxsize=85mm
                \epsfbox[20 370 580 470] {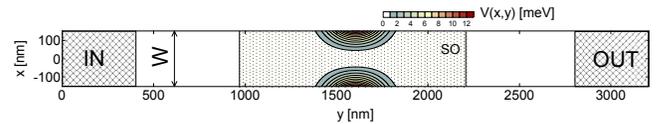}
                 \caption{Sketch of the considered system. The SO interaction is present in the dotted area. The color contours presents the constriction for $\hbar\omega=1.24$ meV}
 \label{fig0}
\end{figure}

The general form of the Rashba Hamiltonian $H_{\mathrm{SIA3D}}=\alpha \nabla V \cdot (\sigma \times \mathbf{k})$ for the two-dimensional system reduces to \cite{nowak}
\begin{equation}
\begin{split}
H_{\mathrm{SIA}} = &\sigma_z\left( \frac{\partial V_{\mathrm{C}}(x,y)}{\partial y}k_x - \frac{\partial V_{\mathrm{C}}(x,y)}{\partial x}k_y\right)\\
-&\frac{i \alpha |e| F_z}{2}  \frac{\partial f(y)}{\partial y}\sigma_x + \alpha|e|  F_z f(y) (\sigma_x k_y - \sigma_y k_x).
\label{hsia}
\end{split}
\end{equation}
Numerical calculations are performed within the model system: a channel of width $W=328$ nm (unless stated otherwise) and length $L=3208$ nm -- see Fig.\ref{fig0},
with $Y_{qpc}=1604$ nm, $l=100$ nm, and the region of SO coupling presence defined by
 $y_1 = 1000$ nm and $y_2 = 2208$ nm -- see the dotted region in Fig. \ref{fig0}. We adopt material parameters for $\mathrm{In}_{0.5}\mathrm{Ga}_{0.5}\mathrm{As}$, i.e.  $m^*=0.0465m_0$, Rashba SO coupling constat $\alpha=0.572\;\mathrm{nm}^2$ after Ref. [\onlinecite{alpha}] and assume $F_z=200$ kV/cm.


We consider electron transport for a fixed Fermi energy $E_f$ for which in a two-dimensional channel the electron's wavevector $k$ takes discrete values. We assume
a small bias between the leads and use the linear Landauer conductance formula $G=\frac{e^2}{h}\sum_i\sum_k T_{i\rightarrow k}$, with summation of transmission probabilities from $i$ subband of the input channel to $k$ subband at the output. For calculation of $T_{i\rightarrow k}$ we solve the Schr\"{o}dinger equation $H\Psi = E\Psi$ on a mesh with $\Delta x = \Delta y = 8$ nm with scattering boundary conditions using the finite difference approach. \cite{szafran}

For the calculation of the spin polarization of the current we impose well-defined spins in the input and output lead. For that purpose we include a weak Zeeman splitting ($E_z \simeq 0.5\;\mu$eV) in the entire
system that corresponds to a residual magnetic field $B=0.1$ mT oriented along the $x$-direction. Next, we separate the conductance to spin positive ($G_+$) and spin negative ($G_-$) currents at the output ($G=G_++G_-$, the formula for $G_+$ contains summation over output modes $k$ that correspond to spin parallel to $x$) and calculate spin polarization of the current as $P=(G_+-G_-)/\lambda$ where $\lambda = (G_++G_-)$ for $G > 2e^2/h$ and $\lambda = 2$ for $G \leq 2e^2/h$.

\begin{figure}[ht!]
\epsfxsize=60mm
                \epsfbox[31 121 572 739] {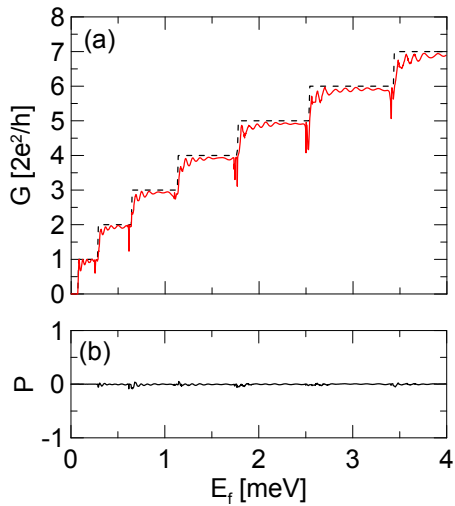}
                 \caption{(a) Conductance versus Fermi energy for a channel without constriction. The dashed black and solid red curves present results obtained without and with the SO coupling in the channel respectively. (b) Mean spin polarization of the current at the end of the channel in the presence of SO interaction.}
 \label{fig2}
\end{figure}

\begin{figure}[ht!]
\epsfxsize=80mm
                \epsfbox[28 651 560 802] {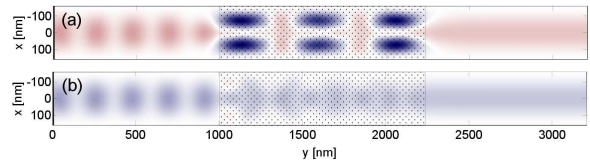}
                 \caption{Spin densities for transport through SO-coupled region marked with squares for $E_f=0.24$ meV. The initial modes have well defined spin oriented along (a) and antiparallel (b) to the $x$-direction. Blue (red) color corresponds to negative (positive) values of the spin density.}
 \label{fig3}
\end{figure}

Let us first assume that the channel is homogenous ($\hbar\omega=0$). The conductance without SO coupling depicted with the dashed curve in Fig. \ref{fig2}(a) has a sharp step-like dependence on the Fermi energy which corresponds to appearance of subsequent subbands at the Fermi level with increasing $E_f$ and the resulting opening of the transport modes each contributing to an increase of conductance by $2e^2/h$. Inclusion of SO interaction in the segment of $y\in(y_1,y_2)$ results in an appearance of conductance oscillations  [the red curve in the Fig. \ref{fig2}(a)] but the main step-like character of the $G(E_f)$ dependence remains unaffected. The spin polarization $P$ of the current is plotted in Fig. \ref{fig2}(b) for the case of active SO coupling. We observe that at the beginning of each conductance plateau the spin polarization exhibits minor oscillations. Besides these oscillations the spin of transmitted electron remains unaffected by the SO coupling and we find $G_+\simeq G_-$.
 Figures \ref{fig3}(a,b) presents spin densities for $E_f=0.24$ meV for which there are two opposite-spin transport modes in the lead. We observe that for the mode where the initial spin is oriented along the $x$-direction [Fig. \ref{fig3}(a)] in the region where SO interaction is active the spin is rotated to -$x$ but after leaving the SO-coupled part it goes back to $x$-positive value. On the other hand the spin negative mode is transferred without spin change [Fig. \ref{fig3}(b)]. As the transmission probability of both the modes are equal, in spite of the spectacular behavior for the incident spin parallel to the $x$ direction, the net spin polarization at the output is zero.

\begin{figure}[ht!]
\epsfxsize=80mm
                \epsfbox[31 180 575 641] {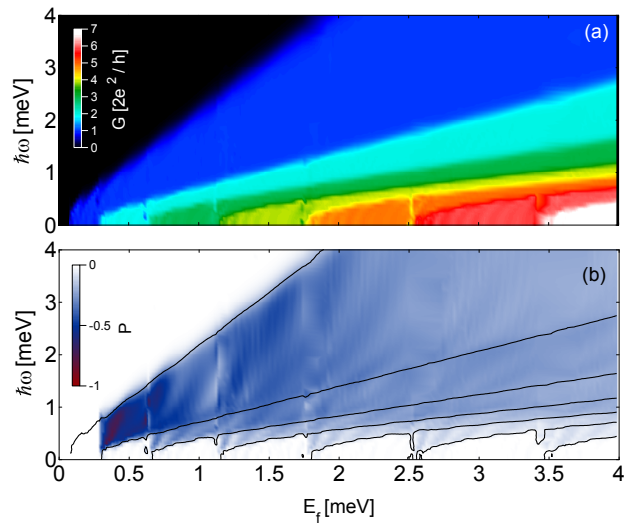}
                 \caption{(a) Map of the conductance as a function of Fermi energy and QPC potential. (b) With colors -- current spin polarization. The curves mark the conductance steps.}
 \label{fig4}
\end{figure}

Let us now introduce QPC in the channel. For a given Fermi energy the constriction limits the number of propagating modes accordingly to its width which results in a step-like dependence of conductance on $\hbar\omega$ depicted in the map of the Fig. \ref{fig4}(a).

\begin{figure}[ht!]
\epsfxsize=65mm
                \epsfbox[34 140 570 700] {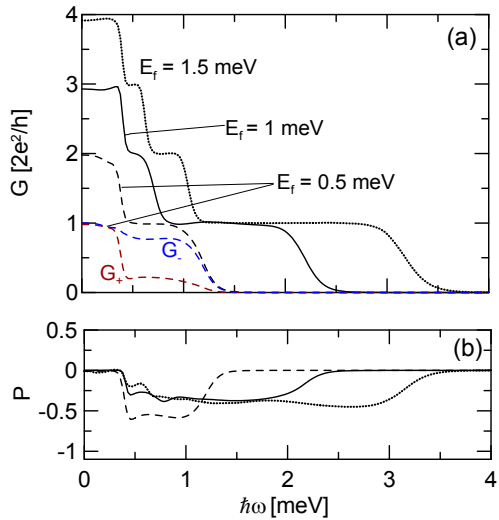}
                 \caption{Cross-sections of Figs. \ref{fig4}(a,b).}
 \label{fig5}
\end{figure}

The map of Fig. \ref{fig4}(b) shows the spin polarization of the output current with colors and the conductance steps with the contours. We observe that at the first conductance plateau for $\hbar\omega=0$ there is no spin polarization in an accordance with Fig. \ref{fig2}(b). Only when the QPC is present the spin polarization appears.\cite{note} The most negative values of $P$ are found on the last plateau for low values of Fermi energy when only two subbands of the leads are conducting. Then, the maximal value of the spin polarization is as large as $P\simeq-0.8$. 
Figures \ref{fig5}(a,b) presents cross-section of maps from Figs. \ref{fig4}(a,b) for three values of Fermi energy. Note that the spin polarization is non-zero at the last conductance step which might be an important issue for the formation of the anomalous plateau.\cite{wang,rejec}

\begin{figure}[ht!]
\epsfxsize=85mm
                \epsfbox[57 174 565 815] {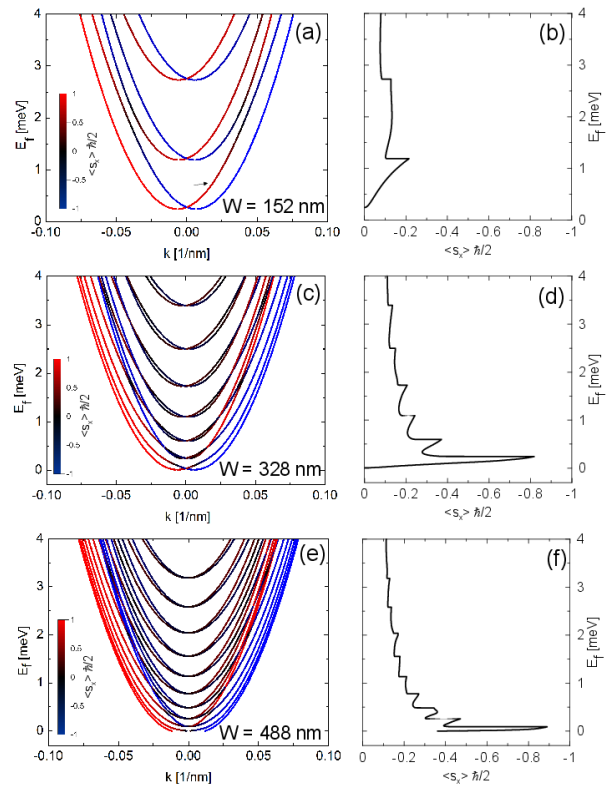}
                 \caption{(a,c,e) Dispersion relation for a channel with the width $W$. The colors of the curves represent mean values of spin $x$ component in each subband. (b,d,f) Mean spin of positive-current modes.}
 \label{fig1}
\end{figure}

In order to understand above finding let us inspect the proprieties of a part of the channel where SO interaction is active.
We consider a straight quantum wire ($\hbar\omega=0$) and plot dispersion relations in Figs. \ref{fig1}(a,c,e) for different channel widths $W$ where the color of the curves depict the mean spin component in the $x$-direction.

For a strictly one-dimensional quantum wire the Rashba coupling exactly polarizes the spins in the direction perpendicular to the channel,\cite{winkler}
 and the dispersion-relation is split into to two parabolas corresponding to the opposite spin orientations.
 Similar result is found for a channel with $W = 152$ nm [Fig.\ref{fig1}(a)] where in the bottom part of the plot ($E_f<0.5$ meV) we find that there are two subbands with almost completely polarized spins.

For a given Fermi energy we calculate the charge current for all the transport modes. Next, we calculate the mean spin polarization of all the transport modes for currents propagating along $y$-direction.
 The result is plotted in Figs. \ref{fig1}(b,d,f). For $W=152$ nm, at low values of Fermi energy ($E_f\simeq0.25$ meV) the opposite spin polarization of positive current modes results in a zero net spin polarization. However as the energy is increased, the mean spin of the positive-current positive-spin subband [marked with arrow in Fig. \ref{fig1}(a)] decreases as a result of spin mixing due to avoided crossing with the higher energy subband. This decrease is translated into an increase of negative net spin polarization of the propagating modes as observed in Fig. \ref{fig1}(b). For $E_f$ for which the second pair of subbands becomes available the net spin polarization abruptly drops.

For wider channel the energy separation between the subbands decreases [see Fig. \ref{fig1}(c) for $W=328$ nm]. This results in an increase of the repulsion between the energy subbands [see the two blue curves at the right bottom part of the dispersion relation in Fig. \ref{fig1}(c)]. Conversely a stronger spin polarization is found for low values of Fermi energy as depicted in Fig. \ref{fig1}(d). The $E_f$ value for which the strongest spin polarization is observed is again just before the appearance of a new set of subbands and the mean spin takes values lower than $-0.8$. However, now the increased spin polarization is observed for narrower range of the energies as compared to the results of Figs. \ref{fig1}(b). The scenario of increasing the net spin polarization as the Fermi energy gets closer to the value where subsequent subbands enter is repeated for higher energies resulting in a saw-like dependence of the mean spin polarization on $E_f$.

For even higher values of the channel width the tendency holds [observe Fig. \ref{fig1}(e,f) for $W=488$ nm] and we note that the spin subbands with negative and positive spin polarization gather forming two bands at higher values of $E_f$ -- see Fig. \ref{fig1}(e).

\begin{figure}[ht!]
\epsfxsize=85mm
                \epsfbox[49 312 570 518] {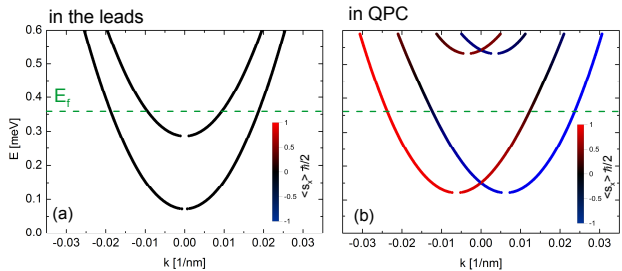}
                 \caption{Dispersion relation calculated in the leads where SO interaction is absent (a) and in the middle of constriction (b). In (a) each band is spin degenerate. Colors of the points depict mean spin in the $x$-direction. The green dashed lines depict Fermi energy used in Fig. \ref{fig6}. Results are obtained for $E_f=0.36$ meV and $\hbar\omega=0.378$ meV.}
 \label{fig7}
\end{figure}

Let us take the parameters for the strong spin-polarization of the current ($P\simeq-0.8$), i.e. $E_f=0.36$ meV and $\hbar\omega=0.378$ meV and plot the dispersion relation in the leads [Fig. \ref{fig7}(a)] -- no SO coupling and width $W=328$ nm and in the channel with lateral potential and SO coupling
that correspond to the center of QPC [Fig. \ref{fig7}(b)]. We find that in latter, there are two modes propagating along the $y$-direction available with mean values of spin operator $\langle s_x\rangle=-0.96\;\hbar/2$ and $\langle s_x\rangle=0.29\;\hbar/2$, respectively -- the net spin polarization is negative similarly to the situation in the low-energy part of plot in Fig. \ref{fig1}(d).\cite{biasso}

\begin{figure}[ht!]
\epsfxsize=85mm
                \epsfbox[35 277 590 812] {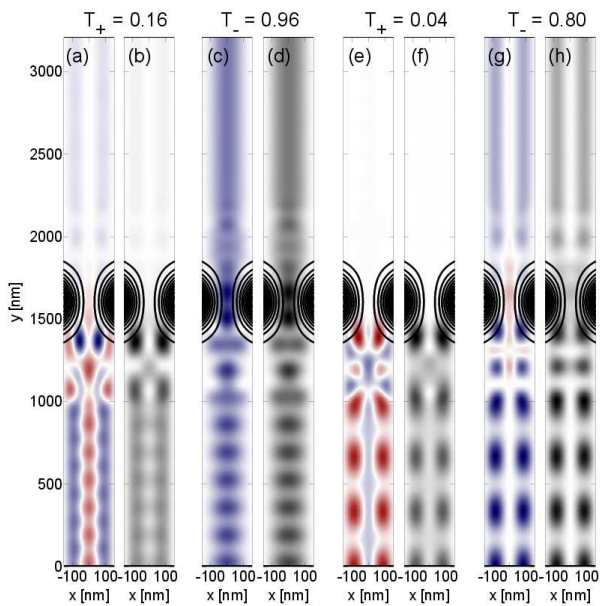}
                 \caption{(a,c,e,g) Spin and charge (b,d,f,h) densities for transport from different lead subbands that appear for $E_f=0.36$ meV and $\hbar\omega=0.378$ meV. Maps (a-d) correspond to $k=0.19$ /nm, and (e-h) to $k=0.095$ /nm -- see Fig. \ref{fig7}. In maps (a,b)(e,f) the incoming electron has the spin polarized along the $x$-direction while in (c,d)(g,h) antiparallel to the $x$-direction. Above the plots calculated transmission probabilities are displayed.}
 \label{fig6}
\end{figure}

The electron incident from the input lead belongs to one of the two nearly-degenerate modes [each of the parabolas in Fig. \ref{fig7}(a) are nearly spin degenerate]. The maps of Figs. \ref{fig6}(b,f) show that the modes where the electron spin is oriented parallel to the $x$-direction [Fig. \ref{fig6}(a,e)] are mainly backscattered at the constriction and the corresponding transfer probabilities are $T_+=0.16$ and $T_+=0.04$. On the other hand the spin negative modes are conducting with $T_-=0.96$ and $T_-=0.80$ [Figs. \ref{fig6}(c,g)]. The limited transport of spin-positive modes from the degenerate pairs of Fig. \ref{fig7}(a) is related to the negative spin polarization of positive-current modes in the QPC for $E_f=0.36$ meV [see Fig. \ref{fig7}(b)] and it is translated into a spin polarization of the current flowing out of the constriction. Moreover we observe that the mode that is initially spin-polarized along the $x$-direction [Fig. \ref{fig6}(a)] after the transfer through QPC gives a contribution to the spin-negative current at the output.

In summary we showed that in a channel with SO interaction there is a spontaneous spin polarization of the positive current modes in the low energy part
of the dispersion relation. The constriction of the QPC induces transient redistribution of the kinetic energy for the incident electron
reducing its wave vector at the transfer across the narrowing.  SO-coupled constriction limits the transfer of the modes with the spin oriented along $x$-direction.
This results in spin polarization of the originally unpolarized electron current.  We find that the polarization is obtained in wide range of Fermi energies and QPC potential with generally being the strongest on the last conductance step [$G\leq2e^2/h$].

This work was supported by the funds of Ministry of Science an Higher Education (MNiSW) for 2012 -- 2013 under Project No. IP2011038671, and by PL-Grid Infrastructure. M.P.N. is supported by the Foundation for Polish Science (FNP) scholarship under START and the MPD Programme co-financed by the EU European Regional Development Fund.


\begin{thebibliography}{00}
\bibitem{note} The results of the spin polarization in the $x$-direction are not affected by the orientation of residual magnetic field. The polarization is obtained also for zero megnetic field.
\bibitem{biasso} The spin poalrization results from SO interaction induced by $F_z$ electric field. The influence of in-plane electric field from the gradient of the QPC potential is found to be negligible here.
\end{thebibliography}
\end{document}